\documentclass[journal]{IEEEtran}
\IEEEoverridecommandlockouts

\usepackage[cmex10]{amsmath}

\usepackage{enumitem}
\usepackage{graphicx}
\usepackage{color}
\usepackage{amsmath}
\usepackage{mathtools}
\usepackage{multicol}
\usepackage{multirow}
\usepackage[english]{babel}
\usepackage{blindtext}
\usepackage{algorithm}
\usepackage{algorithmic}
\usepackage{balance}
\usepackage{amsfonts}
\usepackage{bm}
\usepackage{stfloats}
\usepackage{subfig}
\usepackage{amsthm}
\usepackage{amssymb}
\usepackage{setspace}
\usepackage[nosort]{cite}
\usepackage{CJK}
\usepackage{cite}
\usepackage{caption}
\usepackage{comment}
\usepackage{array,multirow}
\usepackage{graphicx}

\theoremstyle{plain}

\usepackage[table]{xcolor}

\begin{document}

\captionsetup[figure]{labelformat={default},labelsep=period,name={Fig.}}

\title{Toward Agile and Cooperative LEO Satellite Beam-Hopping Networks: Paradigms, Challenges, and Opportunities}

\author{Xinyi Huang,
        Bodong Shang,
        and Meixia Tao
\thanks{Xinyi Huang and Bodong Shang (corresponding author) are with College of Information Science and Technology, Eastern Institute of Technology, Ningbo, Zhejiang 315200, China.}
\thanks{M. Tao is with the School of Information Science and Electronic Engineering, Shanghai Jiao Tong University, Shanghai 200240, China.}
\thanks{This work has been submitted to the IEEE for possible publication. Copyright may be transferred without notice, after which this version may no longer be accessible.}
}

% make the title area
\maketitle

\begin{abstract}
% 1-(50 words, BG)
Low-Earth orbit (LEO) satellite beam-hopping (BH) technology is emerging as a promising approach to meet the ever-increasing global connectivity demands, enabling agile, on-demand coverage.
%benefits
LEO satellite BH can address the spatio-temporal non-uniformity of ground user traffic by dynamically allocating capacity and optimizing network performance. Cooperative multi-satellite BH enables joint transmission and interference avoidance to improve received signal quality.
% 2-(100 words, overview)
% Architecture
This article provides a comprehensive paradigm of BH, detailing its key dimensions, strategies, and architectures.
%Challenge Applications
Through exploration of key challenges, including beam pattern design, on-demand scheduling, and interference management, this paper identifies the potential applications of BH, ranging from adaptive capacity allocation for hotspot areas, low-power Internet-of-Things (IoT), delay-sensitive services, to massive connectivity support.
% system-level analysis
Furthermore, a system-level analysis is presented, including key metrics, models of inter-beam and inter-satellite interference, and cooperative joint transmission, and a case study is provided to demonstrate the performance benefits of BH with cooperative transmission.
%Rest contents
Several promising future research directions are discussed to guide the future development of LEO satellite BH networks.
\end{abstract}

% \begin{IEEEkeywords}
% IEEE, IEEEtran, journal, \LaTeX, paper, template.
% \end{IEEEkeywords}

\IEEEpeerreviewmaketitle

\section{Introduction}
% Para 1: Need for BH
As traditional terrestrial networks struggle to meet the increasing communication demands of remote areas, maritime regions, and disaster-stricken zones, low-Earth orbit (LEO) satellites have emerged as a vital complement, offering low-latency, high-throughput, and ubiquitous coverage \cite{10530195}.
To accommodate the needs of multiple users communicating simultaneously with a satellite, the satellite can transmit multiple beams concurrently using a phased-array antenna. This allows them to communicate with multiple users simultaneously, thereby significantly increasing system capacity and reducing per-user latency. However, ground user demands are non-uniform in both time and space dimensions. 
% Para 2: BH 
One of the most promising technologies within LEO satellite systems is beam-hopping (BH), which allows satellites to dynamically allocate resources, optimizing both time and spatial coverage. BH technology, representing an advancement over traditional fixed multi-beam systems, can effectively address spatio-temporal demand non-uniformity.
LEO satellites employing BH technology can achieve improved system capacity, optimized data throughput, greater flexibility, and superior scalability by enabling dynamic beam switching across multiple cells within their coverage. With rising demand for more efficient and adaptable communication, a thorough understanding of LEO satellites' BH paradigms, technical challenges, typical applications, and system-level analysis is crucial to fully unlock the potential in next-generation satellite networks.

% Para 3: BH Problems + Related works
Some research focuses on the allocation of resources, such as power and bandwidth, alongside BH patterns. 
For multi-beam satellite systems, a hybrid optimization method for BH and power allocation was proposed in \cite{ran2024towards}. In \cite{zheng2024traffic}, the authors optimized the time-frequency-power multi-dimensional resource allocation to address the differentiated quality of service (QoS) requirements.
Other researchers primarily study the optimal design of BH patterns and interference avoidance.
\cite{chen2022next} proposed selective precoding techniques to enhance BH efficiency. Selective precoding allows a subset of beams to be jointly precoded based on their spatial proximity and interference levels.
Furthermore, some studies analyze BH performance using a queueing model. \cite{feng2023performance} evaluated the impact of traffic dynamics on throughput and latency by establishing a discrete-time queueing theory model. \cite{jia2024dynamic} proposed an algorithm based on Lyapunov optimization to minimize system cost while ensuring queue stability.

Despite the growing interest in LEO satellite BH, most existing studies primarily focus on resource allocation and beam scheduling within a single satellite or in non-cooperative multiple-satellite settings. Modern LEO constellations are evolving toward dense multi-satellite deployments. In such scenarios, BH can potentially evolve beyond single-satellite operation to enable cooperative multi-satellite communication, including joint transmission and interference avoidance. However, the opportunities and system-level implications of this cooperative paradigm have not yet been sufficiently discussed.

Therefore, this article aims to provide a comprehensive overview of the LEO satellite BH and to highlight its emerging role in enabling cooperative multi-satellite communication.

% Para 4: Contributions
The main contributions are summarized as follows:
\begin{itemize}
    \item We present a comprehensive overview of BH, introducing its key dimensions, strategies, and architectures. Specifically, we compare temporal and spatial BH, static versus dynamic BH patterns, and centralized, distributed, and cooperative BH architectures.
    
    \item We explore the challenges in LEO satellite BH, including the beam pattern design, on-demand scheduling, and interference management. We further discuss cooperative multi-satellite BH, enabling joint transmission and cooperative interference management in dense LEO constellations.
    
    \item We propose four typical applications for LEO satellite BH, including adaptive capacity allocation for hotspot areas, low-power Internet of Things (IoT), delay-sensitive services, and massive connectivity support.

    \item We conduct a system-level analysis of LEO satellite BH, providing key metrics, inter-beam and inter-satellite interference modeling, and cooperative joint transmission modeling. Then, a case study is presented that illustrates the performance benefits of BH with cooperative transmission.

    \item We also discuss future research directions in LEO satellite BH systems, proposing advancements in BH in satellite-terrestrial integrated networks (STINs), BH-based joint communication and computing design, BH-empowered low-altitude economy, and security- and privacy-preserving BH design.
\end{itemize}.

\section{LEO Satellite Beam-Hopping Paradigms: Dimensions, Strategies, and Architectures}
\begin{figure*}[tp]
	\begin{center}
		\setlength{\abovecaptionskip}{+0.1cm}
		\setlength{\belowcaptionskip}{+0cm}
		\centering
		\includegraphics[width=6.4in, height=3.1in]{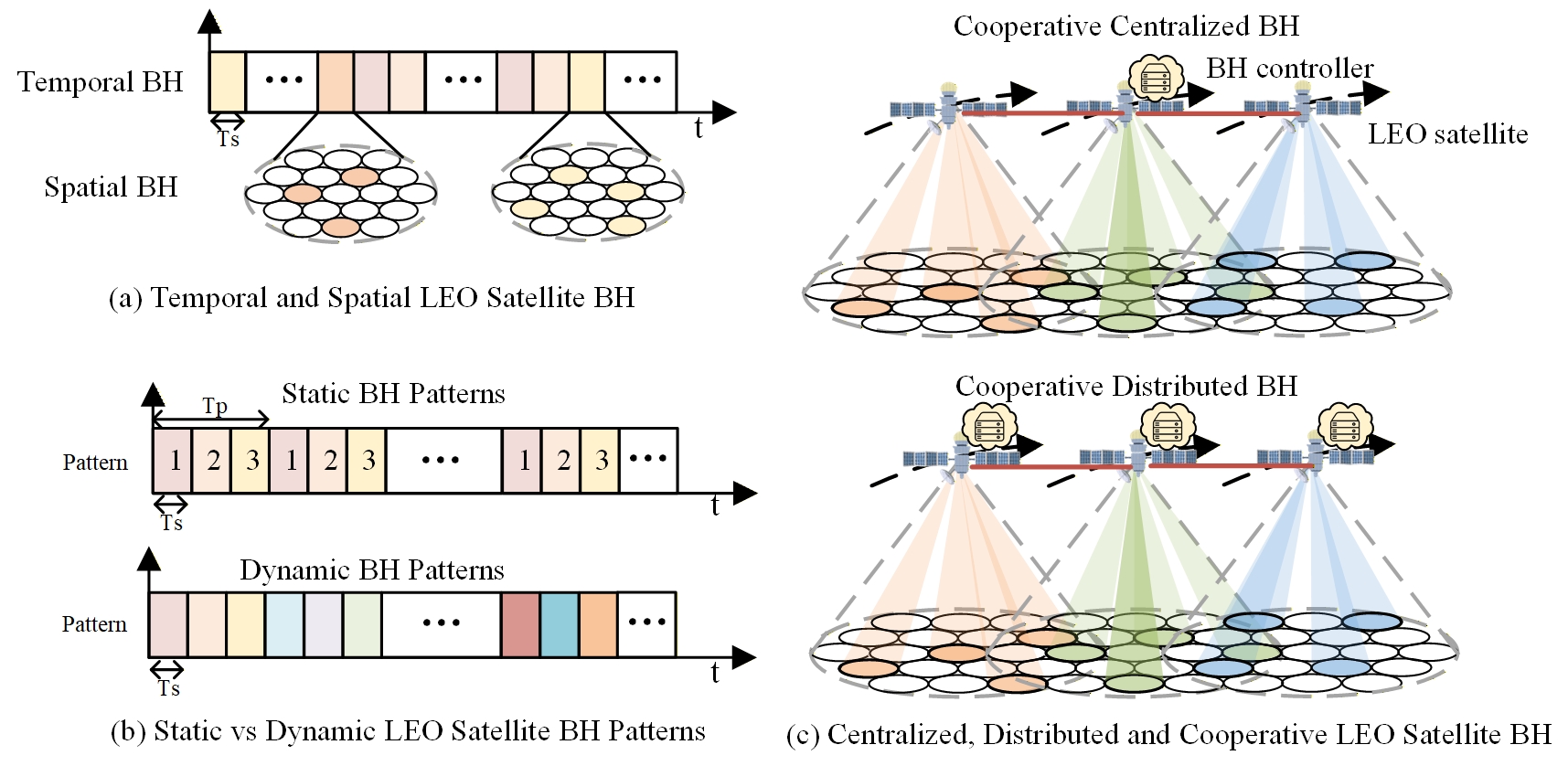}
		\renewcommand\figurename{FIGURE}
		\caption{\scriptsize A sketch of LEO satellite BH.}
		\label{fig:Paradigms}
	\end{center}
	\vspace{-8mm}
\end{figure*}

As illustrated in Fig. \ref{fig:Paradigms}, LEO satellite BH paradigms are investigated from three perspectives, including dimensions, strategies, and architectures.

\captionsetup{font={scriptsize}}

\subsection{Dimensions: Temporal and Spatial LEO Satellite BH}
According to different dimensions of the LEO satellite BH, the BH process can be viewed from two closely coupled perspectives, namely the temporal dimension and the spatial dimension, as illustrated in Fig. \ref{fig:Paradigms} (a).
% In LEO satellite communication systems, BH is a dynamic resource allocation technology that optimizes resource utilization by flexibly controlling the activation time and spatial coverage of beams.
\subsubsection{\textbf{Temporal LEO Satellite BH}}
Time is partitioned into discrete time slots, denoted $T_S$ in Fig. \ref{fig:Paradigms}, with only a subset of beams activated in each slot. The system employs time-slicing technology for beam operation, where beams are alternately activated in turn over time slots to serve distinct cells. Temporal BH focuses on beam activation and revisiting behavior over time to address time-varying traffic demands.
% , where beam activation follows a predetermined scheduling pattern to periodically serve individual cells
\subsubsection{\textbf{Spatial LEO Satellite BH}}
Satellite coverage area is divided into multiple service cells, with each beam serving a specific cell. Spatial BH dynamically adjusts beam pointing, enabling multiple beams to serve different cells simultaneously. 
% This technology is typically implemented through phased-array antennas or other advanced beamforming techniques. The system adopts frequency reuse due to the beam spatial isolation. 
By carefully coordinating beam patterns, the system can reuse the same frequency bands across spatially separated cells, thereby enhancing spectral efficiency. However, such reuse introduces co-channel interference (CCI) that should be mitigated.
Spatial BH describes the allocation of beams across different geographic cells within a given time snapshot, enabling adaptive coverage for non-uniform or mobile users.
\subsubsection{\textbf{Summary}}
Temporal and spatial LEO satellite BH represent two complementary and inherently coupled dimensions of the BH process. 
In practical LEO satellite BH implementations, BH decisions jointly involve temporal scheduling and spatial beam allocation, and the system dynamically coordinates these two dimensions based on real-time ground traffic demands to achieve optimal performance.

% Temporal BH focuses on beam activation and revisiting behavior over time to address time-varying traffic demands, while spatial BH describes the allocation of beams across different geographic cells within a given time snapshot, enabling adaptive coverage for non-uniform or mobile users.

\subsection{Strategies: Static vs Dynamic LEO Satellite BH}
LEO satellite BH employs two fundamental scheduling strategies: static BH patterns with fixed, predetermined resource allocation, and dynamic BH patterns featuring adaptive real-time beam reconfiguration, as shown in Fig. \ref{fig:Paradigms} (b). 
 % for stable traffic conditions  to accommodate fluctuating demand
% These two strategies demonstrate the essential trade-off between implementation simplicity and operational flexibility.
\subsubsection{\textbf{Static LEO Satellite BH}}
Beams operate with predetermined coverage areas and fixed hopping sequences, serving different cells rigidly and periodically, regardless of real-time traffic fluctuations or user distribution. This strategy offers operational simplicity but lacks the adaptability to dynamic network demands, which is suited for periodic and uniform traffic loads.
% and doesn't fluctuate significantly.
\subsubsection{\textbf{Dynamic LEO Satellite BH}}
Dynamic BH intelligently adjusts beam coverage, hopping sequences, and dwell times in real time based on traffic demands, enabling on-demand resource allocation for non-uniform traffic, which is well-suited for regions with uneven and rapidly changing traffic. This flexibility, however, comes at the cost of increased scheduling complexity.
It operates via a closed-loop control framework requiring real-time inputs, such as single-cell traffic statistics, satellite ephemeris and beam footprint prediction, and interference measurements. The signaling overhead primarily stems from reporting traffic state information and distributing updated beam schedules. To mitigate signaling overhead, traffic reports can be quantized or aggregated. 
For LEO satellites moving at $7$-$8$ km/s, the beam dwell time is typically on the millisecond scale. To meet this requirement, a pipelined architecture is commonly employed. Specifically, the scheduler can compute a sequence of BH decisions for several upcoming time slots in advance, while the satellite payload directly executes the precomputed decisions. In practice, the decision computation and execution can be performed in parallel on separate processing units, enabling continuous operation without requiring the scheduling algorithm to finish within a single dwell time.
% While dynamic BH offers high resource utilization, it demands complex scheduling algorithms to compute optimal hopping strategies. 
%. This flexibility is driven by cell distribution and actual traffic demands, ultimately enabling on-demand resource allocation
% , including factors like load and service priority
\subsubsection{\textbf{Summary}}
In practical BH, a hybrid strategy combining a semi-static scheduling framework with a dynamic adjustment mechanism can be employed. Overall, dynamic BH is well-positioned to meet the growing and ever-changing global communication needs.
%, particularly in scenarios involving uneven and time-variable ground traffic distribution.
% This involves dynamically adjusting the activation duration of certain beams within a predetermined static cyclical frame, based on real-time demands.
% is the future trend for LEO satellite communication as it
%

\subsection{Architectures: Centralized, Distributed, and Cooperative LEO Multi-Satellite BH}
% The architecture of LEO satellite BH systems plays a crucial role in determining their responsiveness, scalability, and interference resilience. 
Depending on the control mechanism, LEO satellite BH architectures can be broadly categorized into centralized and distributed implementations, as illustrated in Fig. \ref{fig:Paradigms} (c).
\subsubsection{\textbf{Centralized LEO Multi-Satellite BH}}
A designated master satellite acts as a central controller to generate BH schedules, collecting global information such as cell demand, satellite positions, and interference, and then optimizing beam activation patterns accordingly. These BH schedules are then transmitted to the respective LEO satellites via inter-satellite links (ISLs).
Although it can potentially achieve global optimality, this approach demands reliable, low-latency control links and introduces signaling overhead, thereby increasing vulnerability.
% This dependency introduces additional signaling overhead and increases vulnerability to single points of failure, especially in highly dynamic or disrupted environments.
\subsubsection{\textbf{Distributed LEO Multi-Satellite BH}}
Each satellite determines its own BH strategy based on locally available information, such as local traffic demand and limited communication with neighboring satellites via ISLs. 
Distributed BH offers enhanced adaptability, robustness, and scalability in a dynamic and growing LEO satellite constellation. However, the lack of a global view may lead to localized optimization, in which individual satellites optimize regional performance at the expense of overall network performance.
\subsubsection{\textbf{Cooperative LEO Multi-Satellite BH}}
Cooperative multi-satellite BH leverages ISLs to coordinate beam patterns across mega-constellations. By jointly optimizing spatio-temporal resources, neighboring satellites can implement coordinated beam-activation schedules to manage inter-satellite interference or perform joint transmission to enhance signal reliability and capacity.
The constellation-level cooperation extends the traditional single-satellite BH paradigm toward a network-centric design, enabling joint transmission, interference management, and spatial diversity gain \cite{shang2026multi}.
\subsubsection{\textbf{Summary}}
As LEO constellations scale up, hybrid architectures combining centralized planning with distributed real-time adjustments and cooperative multi-satellite BH are emerging as promising solutions, aiming to balance global efficiency and local flexibility.

\section{Challenges and Solutions in LEO Satellite Beam-Hopping}
This section categorizes and addresses the main challenges faced in LEO satellite BH, along with recent research and practical techniques proposed to tackle these challenges.
\subsection{Beam Pattern Design}
%challenge
Designing efficient and flexible beam patterns for LEO satellites is challenging due to rapid satellite motion, diverse user distribution, and dynamic service demands.
A key issue is determining the optimal number of activated beams to balance coverage, QoS, and system capacity. 

From a system-level perspective, stochastic geometry (SG) provides a tractable analytical framework for modeling and evaluating coverage probability and system capacity as functions of beam and satellite densities. By deriving performance trends, SG enables designers to identify performance bounds before implementing the algorithm.
At the algorithmic level, optimization and learning-based methods realize adaptive beam control. 
Convex or non-convex optimization for beam activation and power allocation aims to maximize throughput under constraints \cite{zheng2024traffic}, but its reliance on channel state information can lead to high computational overhead. Deep reinforcement learning (DRL) models beam scheduling as a Markov decision process \cite{wang2024resource}, in which satellites observe traffic states and iteratively learn policies to optimize long-term objectives in dynamic environments, though this approach requires significant training. Heuristic algorithms, such as genetic algorithms (GA) \cite{deng2024satellites}, search for near-optimal beam activation patterns by iteratively evolving candidate solutions. Though less dependent on precise modeling assumptions, the scalability may be limited.

Overall, SG-based analysis provides performance bounds, while optimization methods enable real-time adaptation, forming a complementary design framework for BH systems.

% Excessive beams increases inter-beam interference, while insufficient beams lead to coverage gaps.

\captionsetup{font={scriptsize}}
\begin{figure}[tp]
\begin{center}
\setlength{\abovecaptionskip}{-0.3cm}
\setlength{\belowcaptionskip}{-0.0cm}
\centering
  \includegraphics[width=3.49in, height=1.9in]{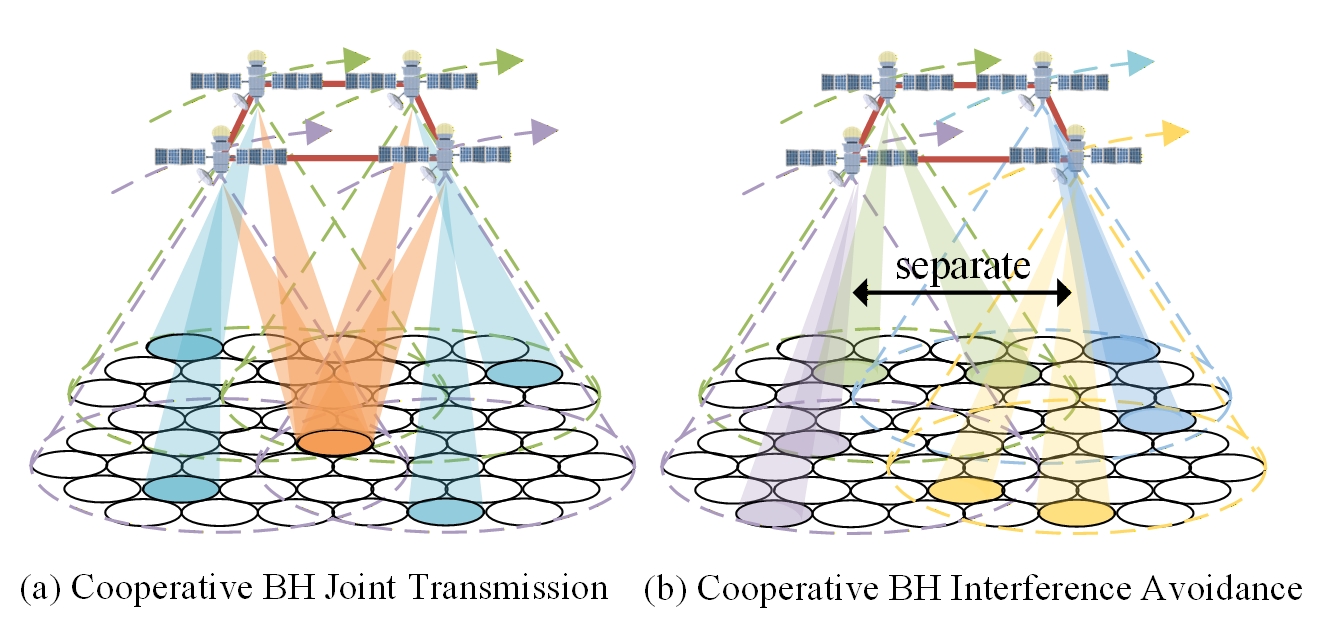}
\renewcommand{\figurename}{FIGURE}
\caption{\scriptsize An illustration of cooperative BH.}
\label{fig:cooperateBH}
\end{center}
\vspace{-7mm}
\end{figure}

\subsection{On-Demand Scheduling and Joint Transmission}
%challenge
On-demand scheduling enables LEO satellite BH to adapt to heterogeneous traffic by dynamically determining beam activation, dwell times, and power allocation. In dense constellations, this mechanism extends from a single satellite to a cooperative multi-satellite framework. Multiple satellites can perform joint transmission by simultaneously illuminating the same cells, enhancing link robustness and capacity, as illustrated in Fig. \ref{fig:cooperateBH} (a).

% solution
Depending on synchronization capabilities and system complexity, joint transmission can operate in different modes. With tight synchronization and shared channel knowledge, satellites operate coherent transmission for maximum capacity. Alternatively, satellites may transmit independently, relying on diversity gain from multiple links with non-coherent transmission. The former offers higher potential combining gain but requires stronger coordination, whereas the latter provides improved robustness with reduced control overhead.
Integrating joint transmission expands BH scheduling from simple beam activation to cooperative alignment across multiple satellites. The scheduler optimizes the satellite selection, temporal window synchronization, and power distribution based on real-time traffic and visibility.
This evolution facilitates targeted capacity reinforcement and gains in spatial diversity, providing a scalable pathway toward multi-satellite cooperative operation in future LEO networks.
%  operates through either coherent combining for maximum capacity or non-coherent diversity for robust connectivity with reduced overhead. 

\subsection{Interference Management in Mega Constellation}
In dense LEO constellations, interference is no longer confined to intra-satellite beam overlap but also arises from simultaneous illumination overlapping or adjacent cells using reused spectrum resources.
In conventional interference mitigation strategies, each satellite independently optimizes its beam pattern, power allocation, or precoding strategy \cite{chen2022next}.
By leveraging inter-satellite coordination, the system can implement spatio-temporal interference avoidance, where neighboring satellites synchronize their hopping sequences to ensure that overlapping footprints never share the same frequency-time resource block, as shown in Fig. \ref{fig:cooperateBH} (b). 
SG provides a tractable framework for modeling the distributions of cells and LEO satellites using binomial point processes (BPP) or Poisson point processes (PPP) \cite{park2022tractable}. This powerful analytical tool helps determine the average interference power a ground cell receives from neighboring satellites in BH, guiding constellation design and enabling optimization of parameters such as satellite orbital altitude and inclination to minimize inter-satellite interference.
At the resource-allocation level, interference-aware scheduling incorporates interference metrics into beam activation decisions, although it relies on timely interference estimation \cite{lin2022multi}. Optimizing power allocation and beam shaping minimizes signal leakage into adjacent beams \cite{tang2021optimization}. Learning-based approaches, such as DRL, learn strategies directly from traffic and interference dynamics, making them suitable for highly non-stationary environments, at the cost of training overhead and limited interpretability \cite{lin2022dynamic}.

\section{Typical Applications of LEO Satellite Beam-Hopping}
\captionsetup{font={scriptsize}}
\begin{figure*}[tp]
	\begin{center}
		\setlength{\abovecaptionskip}{+0.1cm}
		\setlength{\belowcaptionskip}{+0.1cm}
		\centering
		\includegraphics[width=6.7in, height=2.3in]{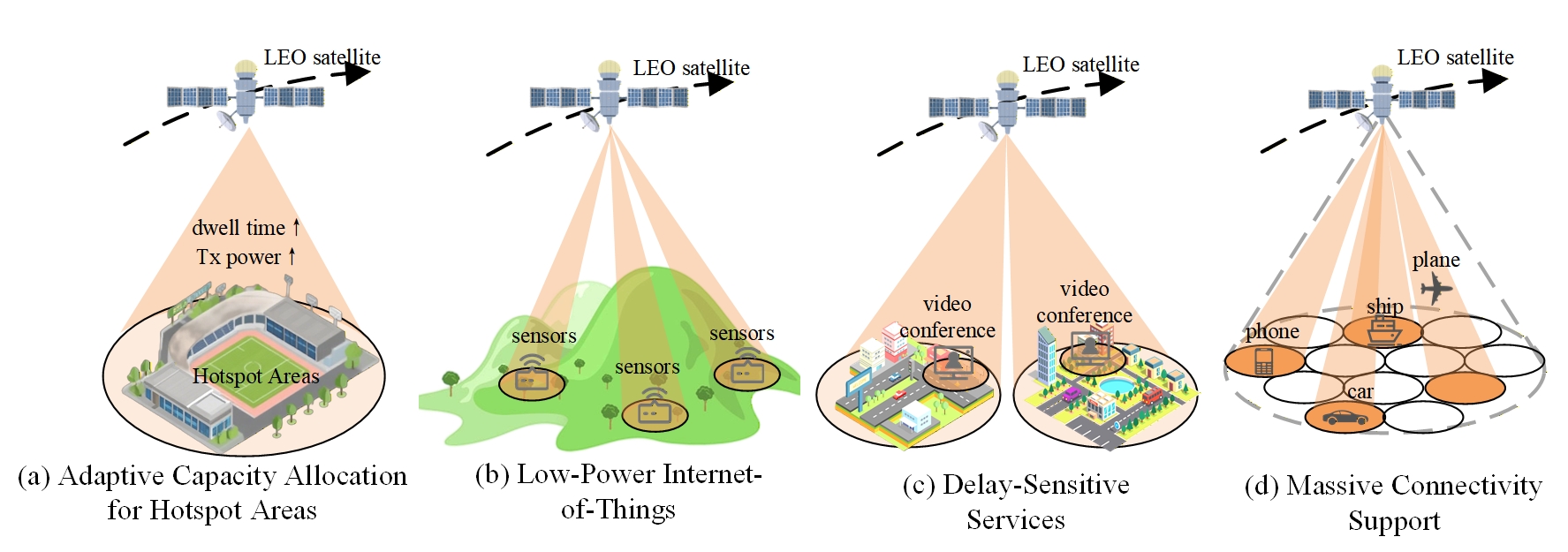}
		\renewcommand\figurename{FIGURE}
		\caption{\scriptsize An illustration of LEO satellite BH applications.}
		\label{fig:application}
	\end{center}
	\vspace{-8mm}
\end{figure*}
This section highlights typical applications of LEO satellite BH, as shown in Fig. \ref{fig:application}, including adaptive capacity allocation for hotspot areas, low-power IoT connectivity, delay-sensitive services, and massive connectivity support.

\subsection{Adaptive Capacity Allocation for Hotspot Areas}
User traffic is often non-uniform and dynamic, creating temporary hotspot areas, such as public events, disaster zones, or congested urban centers. LEO satellite BH is uniquely suited for these scenarios.
By dynamically adjusting beam dwell time, transmit power, or beam activation density, BH concentrates capacity on demand bursts, as shown in Fig. \ref{fig:application} (a). This adaptive approach ensures high-quality service and prevents congestion or interruptions. Unlike traditional fixed multi-beam systems, BH directs resources to follow traffic, eliminating waste in low-demand areas and significantly enhancing overall system resource utilization.

\subsection{Low-Power Internet-of-Things}
Massive IoT deployments, particularly sensors in remote, maritime, or airborne environments, often lack adequate terrestrial connectivity. As shown in Fig. \ref{fig:application} (b), LEO satellite BH provides an essential wide-area complement.
Satellite-enabled IoT terminals are typically power-constrained and characterized by intermittent, low-rate transmissions of small data packets. By dynamically adjusting beam coverage and flexibly scanning wide areas, BH efficiently collects sporadic data from dispersed devices. BH integrates with traffic shaping by aggregating these packets into specific hopping windows that are synchronized with device wake cycles. From a queueing perspective, IoT packets can be buffered until the next beam dwell period, and the revisit interval effectively determines the maximum waiting time. Consequently, BH enables a service paradigm tailored to intermittent IoT traffic through efficient buffer management and periodic resource allocation.
% n energy-efficient

\subsection{Delay-Sensitive Services}
Delay-sensitive applications, such as video conferencing, interactive navigation, and emergency communications, demand consistent and predictable end-to-end latency. As illustrated in Fig. \ref{fig:application} (c), BH reduces communication latency across various scenarios by ensuring rapid beam response and connection establishment. 
Although propagation delay dominates the physical latency of LEO satellite links, BH can still reduce the service latency users experience. Specifically, LEO satellite BH enables delay-aware resource scheduling, providing deterministic beam revisit intervals and minimizing queueing delays, both essential for these services.
Unlike the periodic aggregation used in IoT, delay-sensitive services are mapped to high-priority queues with shorter hop cycles. Specifically, BH prioritizes delay-sensitive traffic within specific time slots, effectively minimizing packet accumulation in satellite buffers and guaranteeing timely coverage for time-critical traffic. BH enables adaptive temporal resource partitioning that dynamically adjusts to heterogeneous QoS requirements and real-time traffic.
% This targeted scheduling effectively minimizes end-to-end delay, ensuring high-quality service for time-critical applications.

\subsection{Massive Connectivity Support}
In the 6G era, supporting massive connectivity remains a challenge for terrestrial networks, particularly in remote, maritime, and airborne environments \cite{li2025advancing}. LEO satellite BH addresses the limitation through efficient spatial and temporal multiplexing.
As shown in Fig. \ref{fig:application} (d), BH enables multi-user and multi-beam reuse by dynamically serving different cells within a service area over time. Furthermore, to accommodate heterogeneous devices with diverse QoS requirements, BH provides a flexible resource allocation mechanism. By adapting to the number and type of connected terminals, the system ensures appropriate connectivity services across varied deployment scenarios.

\section{System-Level Analysis for LEO Satellite Beam-Hopping}

System-level evaluation of BH in LEO satellite networks is essential for understanding performance trade-offs.

\subsection{Key metrics}
Comprehensive evaluation of BH systems involves a multi-dimensional set of metrics that capture network performance from both cell-centric and system-wide perspectives.
Coverage probability in LEO satellite BH refers to the probability that a user is within a satellite's service area. Mathematically, coverage probability is the probability that a user meets a signal-to-interference-plus-noise ratio (SINR) threshold.
Moreover, achievable data rate refers to the amount of data a user or area can transmit within a time slot. Drawing from the principles of the Shannon-Hartley theorem, the achievable data rate is fundamentally determined by the bandwidth and the SINR. 
In addition, system capacity refers to the total throughput a system can support under specific conditions. Mathematically, system capacity is often expressed as the sum of achievable data rates across all active beams, indicating the total throughput over these beams.
% , a metric directly influenced by the BH pattern

\subsection{Inter-Beam and Inter-Satellite Interference Modeling}
Interference in LEO satellite BH systems can be decomposed into inter-beam interference and inter-satellite interference. A tractable system-level analysis requires explicit statistical modeling of both components.

\textbf{Inter-beam interference:} Multiple beams are activated simultaneously within a single satellite to serve adjacent cells, resulting in CCI due to aggressive frequency reuse. The interference depends on three key factors: angular separation between beams, determining antenna gain overlap and side-lobe leakage; transmit power allocation across active beams; and propagation loss determined by satellite altitude and user location. In practice, antenna radiation patterns are mapped to gain-versus-angle profiles, and the aggregate inter-beam interference is obtained by summing the received power from activated beams except the serving beam.
%  This modeling explicitly captures the trade-off between spatial reuse and interference escalation as the number of active beams increases.
% A larger angular separation generally results in less interference, while a smaller angle leads to more significant interference.

\textbf{Inter-satellite interference:} In mega-constellations, satellites are spatially distributed over the orbital sphere. By modeling satellite locations as a homogeneous PPP with density $\lambda_S$ on the spherical surface, the aggregate inter-satellite interference at a typical user can be statistically modeled by integrating the transmit power, path loss, and antenna gains across all visible satellites. Accurately modeling inter-satellite interference requires a thorough analysis of the probability that two satellites simultaneously illuminate shared or adjacent ground regions. Using tools from SG \cite{park2022tractable}, such as the probability generating functional of a PPP, the Laplace transform of the aggregate interference can be derived, which enables analytical characterization of the SINR distribution. This approach can evaluate coverage probability, achievable rate, and system capacity without relying on deterministic constellation layouts.

By combining these two components, we can derive closed-form or semi-analytical expressions for key performance metrics. Such system-level modeling provides performance bounds and design insights independent of algorithms, forming the analytical foundation for optimization strategies.

\subsection{Cooperative Joint Transmission Modeling}
Cooperative joint transmission provides an analytical framework for LEO multi-satellite BH systems in which multiple satellites jointly serve the same ground terminal. In this paradigm, the received signal is no longer characterized by a single desired link and independent interference terms. Instead, it is modeled as the coherent or non-coherent superposition of synchronized signals from a cluster of coordinating satellites \cite{shang2026multi}. This shift fundamentally redefines the SINR calculation by converting potential inter-satellite interference into useful signal power, while dynamically filtering the remaining interference pool to exclude satellites within the cooperative set.
% Instead, it is modeled as the superposition of signals transmitted by a cluster of coordinating satellites, while the remaining satellites constitute the interference pool.

The modeling process consists of three key steps. First, a cooperation cluster is defined based on criteria, such as satellite visibility, angular proximity, or channel quality. Second, the received signal power is obtained by aggregating signals from satellites within the cluster, which may be combined coherently or non-coherently depending on the synchronization capability and coordination overhead. Third, the interference field is reconstructed by excluding the cooperating satellites and statistically characterizing the remaining in the constellation.

By integrating these cooperative factors, the analysis provides a tractable evaluation of performance gains brought by cooperative transmission.

\begin{figure*}[t]
\centering
\renewcommand{\figurename}{FIGURE}
\setlength{\abovecaptionskip}{+0.2cm}
\subfloat[Achievable data rate for non-/cooperative BH]
{\includegraphics[width=2.33in, height=1.9in]{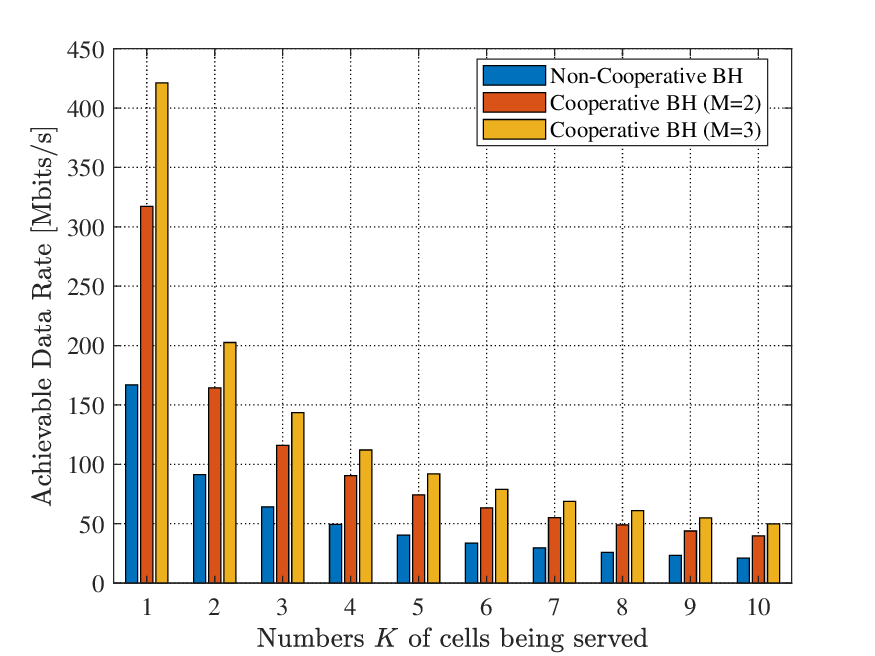}
\label{fig4a}}
\subfloat[System capacity for non-/cooperative BH]
{\includegraphics[width=2.33in, height=1.9in]{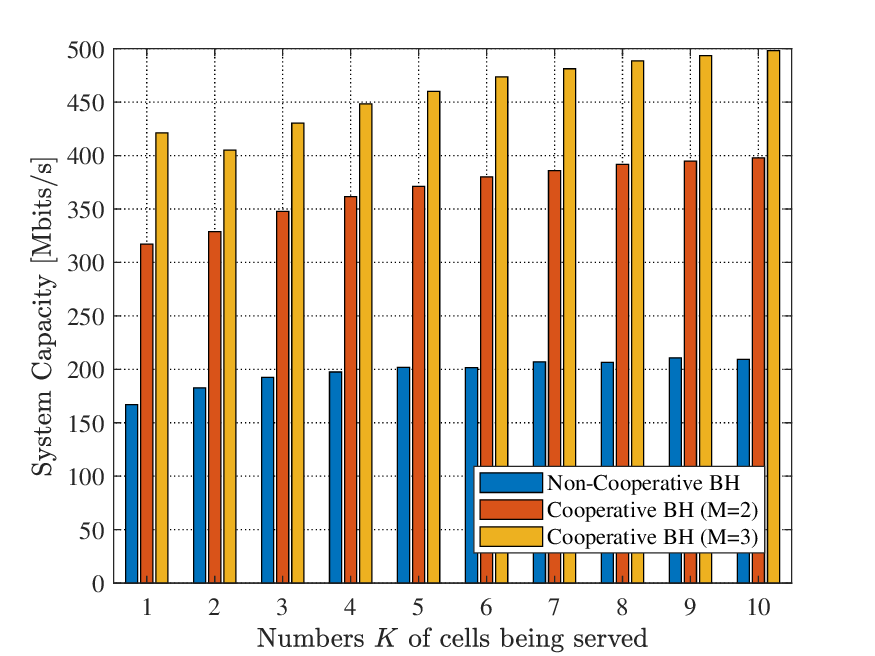}
\label{fig4b}}
\subfloat[Average delay for non-/cooperative BH]
{\includegraphics[width=2.33in, height=1.9in]{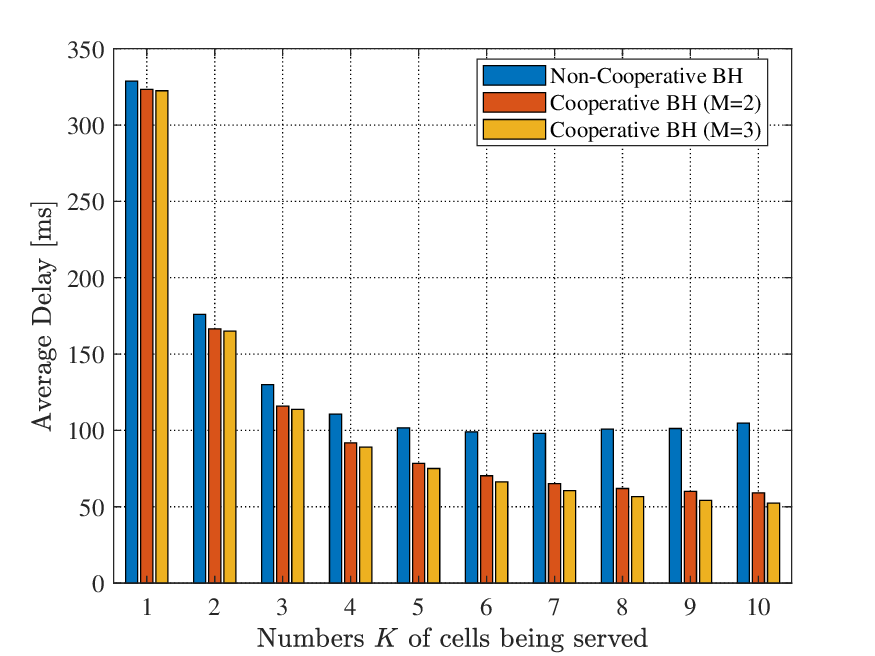}
\label{fig4c}}
\caption{Performance comparison between non-cooperative and cooperative BH.}
\label{fig:casestudy}
\vspace{-5mm}
\end{figure*}

\section{Case Study: Cooperative LEO Satellite Beam-Hopping Performance}
% Cooperative and Non-Cooperative Beam-Hopping for Data Rate and Capacity
To further compare non-cooperative and cooperative LEO satellite BH performance, we conduct simulations of the achievable data rate within a time slot, the system capacity during a specific period and average delay under different numbers of cells simultaneously being served, enabling a detailed case study, as shown in Fig. \ref{fig:casestudy}.
In this case, LEO satellites are distributed according to PPP with a density of $\lambda_S=1\times 10^{-6}$/km$^2$ and a height of $500$km with $128$ service cells area. The transmit power of $P_T=20$dBW is set for all active BH beams, and the antenna gain given by 3GPP \cite{3gpp.38.863}, with system bandwidth $B=100$MHz and the time slot $5$ms \cite{tang2021optimization}. The signal propagation is modeled by path-loss attenuation with the path loss exponent $\alpha=2$ and small-scale fading by the $m=2$ Nakagami-m distribution \cite{park2022tractable}.

In Fig. \ref{fig:casestudy} (a), the achievable data rate decreases monotonically with different numbers $K$ of cells are simultaneously served in BH, due to the intensified inter-beam interference within a time slot. However, cooperative BH significantly outperforms non-cooperative BH across all $K$ values, with the improvement becoming more pronounced as the number of cooperating satellites $M$ increases. 
In contrast, Fig. \ref{fig:casestudy} (b) shows that the system capacity increases with $K$. Cooperative BH achieves higher capacity than non-cooperative BH, and the gain grows with $K$. This highlights the capability of cooperative strategies to scale efficiently with network density.
Fig. \ref{fig:casestudy} (c) presents the average delay performance. Delay decreases with $K$ due to reduced queueing time, and cooperative BH maintains a lower delay than non-cooperative BH.
Overall, the results demonstrate that five cells offer a favorable trade-off among data rate, capacity, and delay, while cooperative BH consistently enhances all performance metrics. At this point, the system capacity approaches saturation, while the achievable data rate continues to decline as $K$ increases further. Alternatively, we can determine the number of cells served by setting a lower bound for the achievable data rate. This also validates the necessity of multi-satellite BH cooperation in dense LEO constellations.

\section{Future Research Directions of LEO Satellite Beam-Hopping}

LEO satellite BH still holds considerable potential to meet the emerging demands in future networks. Several important future research directions are explored, as shown in Fig. \ref{fig:future}.
\subsubsection{\textbf{BH in STINs}}
LEO satellite BH adapts to increasingly heterogeneous topologies in STINs, where satellites and base stations cooperate to serve cells, as shown in Fig. \ref{fig:future} (a). 
Unlike existing STIN coordination frameworks that assume static satellite beams and continuous coverage, BH introduces time-slotted and spatially discontinuous illumination. This leads to a fundamentally cross-segment coupling problem, in which terrestrial scheduling must be aligned with beam revisit cycles and dwell-time allocation.
% STINs are vital for 6G to provide ubiquitous and seamless connectivity.
% While coordination with terrestrial networks increases complexity, it offers significant optimization opportunities.
\subsubsection{\textbf{BH-Based Joint Communication and Computing Design}}
Integrating LEO satellite BH with joint communication and computing represents a pivotal evolution, as illustrated in Fig. \ref{fig:future}(b). Existing satellite edge computing frameworks typically assume persistent communication links, while, under BH, communication opportunities become time-fragmented. This requires revisit-aware task partitioning, beam-synchronous offloading strategies, and latency-constrained computing-window scheduling. Also, the coupling between beam dwell time and onboard resource allocation constitutes a new cross-layer optimization problem.
\subsubsection{\textbf{BH Empowered Low-Altitude Economy}}
Low-altitude economy, comprising uncrewed aerial vehicles (UAVs) and low-altitude monitoring, requires high mobility, low latency, and reliability. LEO satellite BH is a key enabler for tracking and serving these dynamic users, as shown in Fig. \ref{fig:future}(c).
Unlike static footprint systems, BH-empowered low-altitude systems must adapt to the rapid three-dimensional spatial dynamics of aerial platforms to maintain stable connectivity. This requires agile beamforming and accelerated beam-switching mechanisms within the BH framework to accommodate high-velocity trajectories and fluctuating spatial distributions.
\subsubsection{\textbf{Security-Aware and Privacy-Preserving BH Design}}
\begin{figure*}[tp]
	\begin{center}
		\setlength{\abovecaptionskip}{+0.1cm}
		\setlength{\belowcaptionskip}{+0.1cm}
		\centering
		\includegraphics[width=6.7in, height=1.75in]{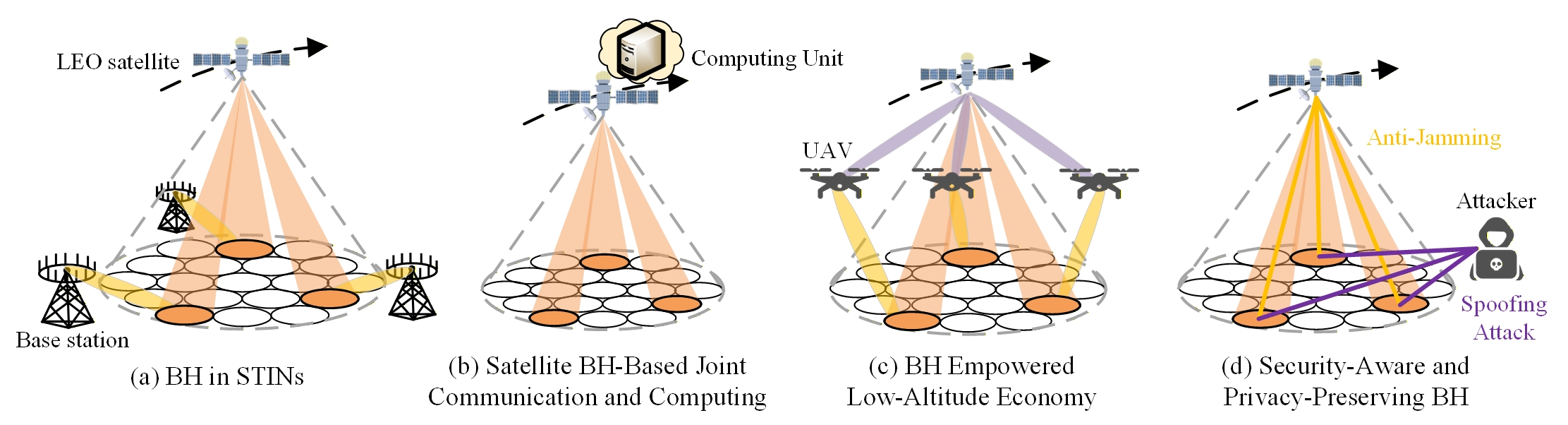}
		\renewcommand\figurename{FIGURE}
		\caption{\scriptsize An illustration of future research directions in LEO satellite BH.}
		\label{fig:future}
	\end{center}
	\vspace{-8mm}
\end{figure*}

As satellite networks integrate with critical infrastructure, security and privacy become paramount. LEO satellite BH introduces unique opportunities and threats. As shown in Fig. \ref{fig:future}(d), cells are subject to spoofing and jamming. To mitigate malicious interference, BH can incorporate anti-jamming strategies such as frequency hopping, adaptive power control, and advanced beamforming. While dynamic beam activation enables physical layer security through pattern obfuscation, poorly secured BH schedules may leak user locations or traffic characteristics.
%  Consequently, researching BH-based protocols for secure and private communication is critical to safeguarding satellite-terrestrial integration.
% , necessitating the integration of anti-jamming and security-enhancing techniques in BH

\captionsetup{font={scriptsize}}

\section{Conclusions}
In this article, LEO satellite BH represents a significant leap forward in satellite communication systems, providing agile, efficient on-demand coverage. We offered insights into its key dimensions, strategies, and architectural paradigms. We investigated the challenges in BH and highlighted cooperative multi-satellite BH in dense LEO constellations by enabling joint transmission and interference management. Furthermore, we presented four representative applications of LEO satellite BH, demonstrating its utility across various domains, and conducted a system-level analysis, including a case study of the performance of cooperative BH transmission. 
Finally, we explored four future research directions, which are essential to unlocking the full potential of the LEO satellite BH. As the LEO constellation continues to expand, BH will play a pivotal role in future satellite communication networks.

\bibliographystyle{IEEEtran}
\bibliography{references.bib}

\section*{Biographies}

\vspace{-12mm}

\begin{IEEEbiographynophoto}{Xinyi Huang}
(xinyihuang@eitech.edu.cn) 
received a B.Eng. degree in communications engineering at East China Normal University, Shanghai, China, in 2025. She is currently pursuing a Ph.D. degree in the joint Ph.D. program between Eastern Institute of Technology (EIT), Ningbo, and the University of Science and Technology of China (USTC), Hefei. Her research interests include space-air-ground integrated networks, satellite MIMO technology, satellite communications, and performance analysis of wireless systems.
\end{IEEEbiographynophoto}

\vspace{-10mm}

\begin{IEEEbiographynophoto}{Bodong Shang}
(bdshang@eitech.edu.cn) 
received his Ph.D. degree from the Department of Electrical and Computer Engineering at Virginia Tech, Blacksburg, USA. 
He is currently an Assistant Professor at the College of Information Science and Technology, Eastern Institute of Technology (EIT), Ningbo, China. 
His research areas are wireless communications and networking, including space-air-ground-sea integrated networks, non-terrestrial networks, and space information networks.
\end{IEEEbiographynophoto}

\vspace{-10mm}

\begin{IEEEbiographynophoto}{Meixia Tao}
(mxtao@sjtu.edu.cn) 
received the Ph.D. degree in electrical and electronic engineering from the Hong Kong University of Science and Technology in 2003. She is a Distinguished Professor with the Department of Electronic Engineering, Shanghai Jiao Tong University, China. Her current research interests include wireless edge learning, semantic communications, integrated communication-computing-sensing, AI-based channel modeling, and beamforming.
She is a Fellow of IEEE.
\end{IEEEbiographynophoto}

\end{document}